\documentstyle[aps,amsfonts,epsfig]{revtex}

    \def\z{\noindent}  
 
    \def\sqr#1#2{{\vcenter{\vbox{\hrule height .#2pt
                             \hbox{\vrule width .#2pt height#1pt \kern#1pt
                                   \vrule width .#2pt}
                             \hrule height .#2pt}}}}

\begin{document}
\wideabs{ \title{Space charge limited 2-d electron flow between two 
flat electrodes in a strong magnetic field}

\author{A.Rokhlenko and J. L. Lebowitz\footnotemark }
\address{Department of Mathematics,
Rutgers University\\ Piscataway, NJ 08854-8019}

\maketitle
\begin{abstract}
            
An approximate analytic solution is constructed for the 2-d space
charge limited emission by a cathode  
surrounded by non emitting conducting ledges of width $\Lambda$. 
An essentially exact solution (via conformal mapping)
of the electrostatic problem in vacuum is matched to the solution
of a linearized problem in the space charge region whose boundaries
are sharp due to the presence of a strong magnetic field.
The current density growth  in a narrow 
interval near the edges of the cathode depends
strongly on $\Lambda$. We obtain an empirical formula for
the total current as a function of $\Lambda$ which extends to
more general cathode geometries.

\medskip

\z PACS: 52.27.Jt; 52.59.Sa; 52.59.Wd; 85.45.Bz

\end{abstract}}
   
\narrowtext

The study of space charge limited (SCL) current, initiated
in the beginning of the last century [1,2], continues to be of great
current interest [3-9]. These works are important for
the design of high power diodes, techniques of charged 
particles beams, physics of non-neutral plasmas including plasma sheath, 
and other numerous applications. The modelling of SCL ionic flow in 
cylindrical and spherical geometry [3] is also necessary for the
inertial-electrostatic confinement of fusion plasmas.
Unfortunately only the planar 1-d case permits an analytic solution [1,2]
and as pointed out in a recent review [5] even ``the
seeming simple problem of 2-d planar emission
remains unresolved''. This motivated the present work which provides 
a semi-analytical solution for a prototype 2-d model similar
to that studied in [6]. We obtain for the first time, we believe,
a reasonable analytic approximation for the currents at the edge of the
cathode - an important (though usually undesirable) feature of SCL
diodes [6,7]. An extension of our methods should facilitate
dealing with this problem to any desirable accuracy thus providing
an alternative to PIC simulations.

\footnotetext{*Also Department of Physics} 
{\it Model}. We consider the current between two conducting flat 
electrodes where the anode, whose potential is $V$, is an infinite plane 
separated by a distance $D$ from the grounded cathode which is an 
infinitely long strip parallel to the anode. Our assumptions 
are: 1) The cathode upper surface, of width $2A$,
has infinite emissivity while the lower face and the ledges
of widths $\Lambda$ 
do not emit (see Fig.1). 2) A very strong strong magnetic field
perpendicular to the electrodes inhibits the transversal components
of electron velocities [6,8], but almost does not affect the total
current [6,8,9]. 3) The emitted electrons leave the cathode with zero 
velocity [1,2,6]. 

If the cathode is in the ($X,Z$) plane and the magnetic field in the 
$Y$-direction the velocities $v$ of electrons are parallel to the $Y$-axis 
with $mv^2(X,Y)=2eU(X,Y)$, where $U(X,Y)$ represents the potential 
field while $m,e$ are the electron mass and charge. The current density 
$J(X)$, which clearly is $Y$-independent, determines together with 
$v(X,Y)$ the density of electrons. Using the dimensionless variables$$
x={X\over D},\ y={Y\over D},\ a ={A\over D},\ \lambda ={\Lambda\over
D},$$
\vskip-1cm
$$\eqno(1)$$
\vskip-1cm 
$$\phi (x,y)={U(X,Y)\over V},\ j(x)=\sqrt{m\over 2e}{9\pi D^2\over 
V^{3/2}}J(X),$$
the nonlinear Poisson equation for the potential then takes the form$$
{\partial^2\phi\over \partial x^2}+{\partial^2\phi\over 
\partial y^2}=-4\pi\rho (x,y)={4j(x)\over 9\sqrt{\phi(x,y)}}.\eqno(2)$$
The electron density $\rho(x,y)$ and current
$j(x)$ are different from zero only in the shaded rectangle Q of Fig.1 
which shows a two dimensional cross section of our system.
\vskip -3cm
\hskip -1cm \epsfig{file=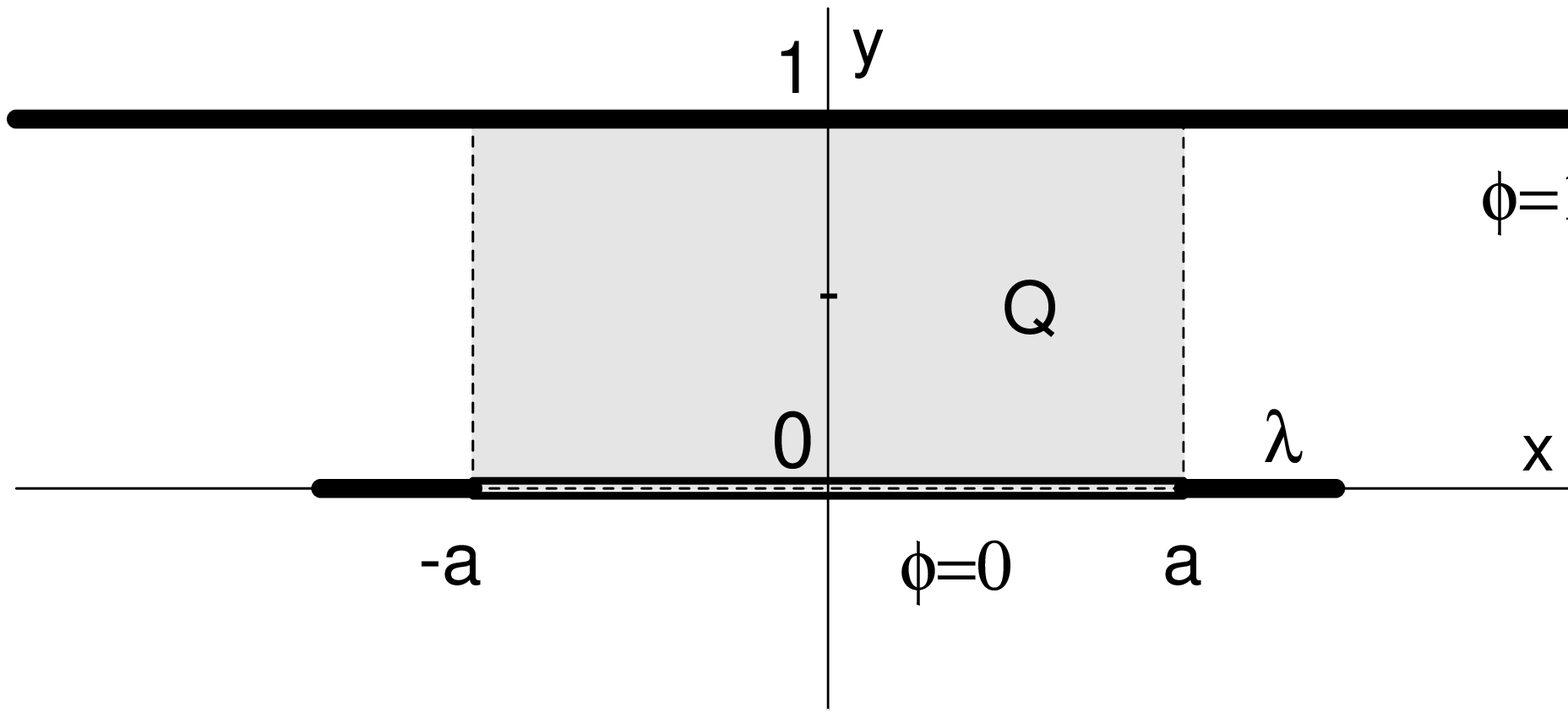, width=8cm,height=8cm}
\vskip-2.5cm
\centerline{\small FIG. 1. Geometry of the system}
\vskip 0.3cm
Eq.(2), subject to the boundary conditions (BC),$$
\phi(x,0)=0\ {\rm for}\ |x|<a+\lambda,\ \phi(x,1)=1\ {\rm for}\ |x|<
\infty ,$$
\vskip-1cm
$$\eqno(3)$$
\vskip-1cm 
$${\partial\phi\over\partial y}(x,+0)=0\ {\rm for}\ |x|<a,$$
is to be solved in the half-plane $y\leq 1$ to produce
both functions $\phi(x,y)$ and $j(x)$ which
are non-negative and symmetric about the $y$-axis.
To do this we first
solve eq.(2) approximately in the current region Q on a rather intuitive
level. The problem is nonlinear here and it is not well posed if
one disregards the field at $|x|>a$. Consequently our 
solution will have a set of free parameters which specify $j(x)$ and
$\phi(x,y)$: in particular $\phi(a,y)$ and ${\partial \phi\over
\partial x}(a^-,y)$. In the second step the potential $\phi(\pm a,y)$ is 
used as the BC and we obtain a Dirichlet problem for 
the Laplace eq.(2) in the outer region of the half-plane where $j(x)=0$.
We solve this problem using conformal mapping techniques and 
evaluate ${\partial \phi\over\partial x}(a^+,y)$. If one excludes 
the points $x=\pm a,\ y=0$ the electron density $\rho(a^-,y)$ 
is finite and $\rho(a^+,y)=0$, therefore the second derivative of
$\phi (x,y)$ has a finite jump at $x=a$, while the first derivative
must be continuous, i.e.$$
{\partial \phi\over\partial x}(a^-,y)={\partial \phi\over\partial x}(a^+,y),
\ 0<y<1.\eqno(4)$$
In the last step we satisfy approximately the matching condition (4)
by adjusting the free parameters mentioned above using the least squares
technique. This will give an approximate explicit form for $j(x)$.

{\it The space charge region} Q. We want to solve approximately eq.(2) 
where the function $j(x)$ is not known nor are the
BC for $\phi$ at $x=\pm a$. 
When $a=\infty$ we have no $x$ dependence 
and (2) becomes an ordinary equation which was solved in [1,2]
yielding $\phi_1 (y)=y^{4/3},\ j_1(x)=1$. This gives the 
Child-Langmuir formula [1], $J_1=(2e/m)^{1/2}V^{3/2}/9\pi D^2$.
{}For $a\gg 1$ it is reasonable to assume that $j(x)\sim j_1=1$  when
$a-|x|\gg 1$ and use also a stronger assumption that the difference
$\phi(x,y)/\phi_1 (y) -1$ is small almost everywhere (i.e. it does not exceed
$\sim 1-1.5$ even near the edges of region). 
This difference however is not small at the cathode edges, $x=\pm a$,
where the electric field must match the field outside. The large 
gradients in the field lead to the acceleration of electrons and thus
to a strong rise of the current density $j(x)$ near the boundary
of the SCL flow.

We represent $\phi(x,y)$ in the form $y^{4/3}[1+\mu(x,y)]$
and linearize the square root as $
[1+\mu(x,y)]^{-1/2}\approx 1-\beta\mu(x,y),$
where the number $\beta$ is chosen to minimize the integral of
$[1-\beta\mu -(1+\mu)^{-1/2}]^2$ on the interval $0\leq \mu\leq 1$. This
yields $\beta \approx 0.328$ with relative average error of approximation
around $2.2\%$. {}For $\mu =0.2,\ 1,\ 1.5$ the error is $2.36\%,\
4.96\%,\ 6.25\%,\ 19.6\%$ respectively. We shall see later that for all
$\lambda \geq 0.1$ $\mu <1.5$. Substituting in (2) we
obtain a linear equation$$
y^2\left ({\partial^2\mu\over \partial x^2}+{\partial^2\mu\over 
\partial y^2}\right )+{8\over 3}y{\partial\mu\over \partial y} 
+4{1+\beta\over 9}\mu ={4\over 9}[j(x)-1],\eqno(5)$$
where we dropped a nonlinear term in the right side. The
error due to this and to the linearization of the square root is 
negligible for small 
$\mu$ and decreases the right side by at most a factor $\sim 0.7$, 
in all the cases considered (see Table 1) including even $\mu\approx 2$.

Using the method of separation of variables we write$$
\mu(x,y)=\sum_lq_lf_l(x)u_l(y),\ j(x)=1+{9\over 4}\sum_l q_lf_l(x),\eqno(6)$$
with$$
f_l(x)=e^{-k_l(a-x)}+e^{-k_l(a+x)},\ |x|\leq a.\eqno(7)$$
Substituting (6) and (7) into (5) and assuming that (5),(3) are
satisfied separately for each $l=1,2,...$ gives a set of 
inhomogeneous equations $$
y^2{d^2u_l\over dy^2}+{8\over 3}y{du_l\over dy}+\left ( k_l^2y^2+
4{1+\beta \over 9}\right ) u_l=1,\eqno(8)$$
with the common BC $u_l(1)=0$. The parameters $k_l$ and $q_l$
will be determined later.  The potential
can be written in the form$$
\phi(x,y)=y^{4/3}+y^{4/3}\sum_l q_lf_l(x)u_l(y),\eqno(9)$$ 
where the first term is the Child-Langmuir potential $\phi_1$ and the
$u_l(y)$ are assumed finite.
The relevant particular solutions of (8), which can be expressed in terms of 
Lommel's functions $s_{-1/6,\nu}(k_ly),\ \nu =\sqrt{9-16\beta}/6$ [10],
is given by the power series expansion
\vskip-0.6cm
$$u_l(y)={9\over 4(1+\beta )}\sum_{n=0}^\infty (-1)^na_n\left ({k_ly\over 2}
\right )^{2n},$$
\vskip-0.9cm
$$\eqno(10)$$
\vskip-0.9cm
$$a_0=1,\ a_n={a_{n-1}\over n^2+5n/6 +(1+\beta )/9}.$$
As all $u_l(1)=0$ the parameters $k_l$ are the increasing
roots of (14): 3.881, 6.675, 10.065, 13.003, 16.316, 19.306, 22.582, 25.600,
28.855, 31.891 for $1\leq l\leq 10$. They can be easily evaluated due to
the rapid convergence of (10), asymptoticaly $k_l\to l\pi$. The 
free parameters $q_l$ will be used to satisfy (4).

The 2-d mean current density over the whole cathode, which in terms 
of our scheme is given by$$
j_2={1\over a}\int_0^aj(x)dx\approx 1+{9\over 4a} 
\sum_l{q_l\over k_l}(1-e^{-2k_l a }),\eqno(11)$$
is usually presented [8] as the 1-d current density $j_1=1$ plus a
correction: $j_2=1+\alpha /2a$. Thus in the original units the mean
current has the form
\vskip-0.6cm
$$J_2=J_1\left (1+\alpha{D\over W}\right ),\eqno(12)$$
where $W=2A$ is the width of the cathode.
Using (11) the parameter $\alpha$ is defined here by$$
\alpha =9\sum_l{q_l\over 2k_l}(1-e^{-2k_l a}).\eqno(13)$$

{\it Electrostatic region}. It seems clear that for $a \gg 1$ 
the electric field in the vicinity of the boundary $x\approx a,\ 0\leq
y\leq 1$
 
is not affected much by the region $x\leq -a,\ 0\leq y\leq 1$, see
Fig.1. This 
 
allows us to study a simpler electrostatic 
problem for a plane which is split according to Fig.2(a). We modified a
conformal transform in [11] to the form
\vskip-0.5cm
$$z=2\pi^{-1}[\ln (\sqrt{w}+\sqrt{w-1})-c\sqrt{w^2-w}],\eqno(14)$$
which maps the shaded half-plane $z=x+iy$ on Fig.2(a) onto 
the upper half-plane $w=u+iv$ in Fig.2(b). 
\vskip -1.3cm
\hskip -0.5cm \epsfig{file=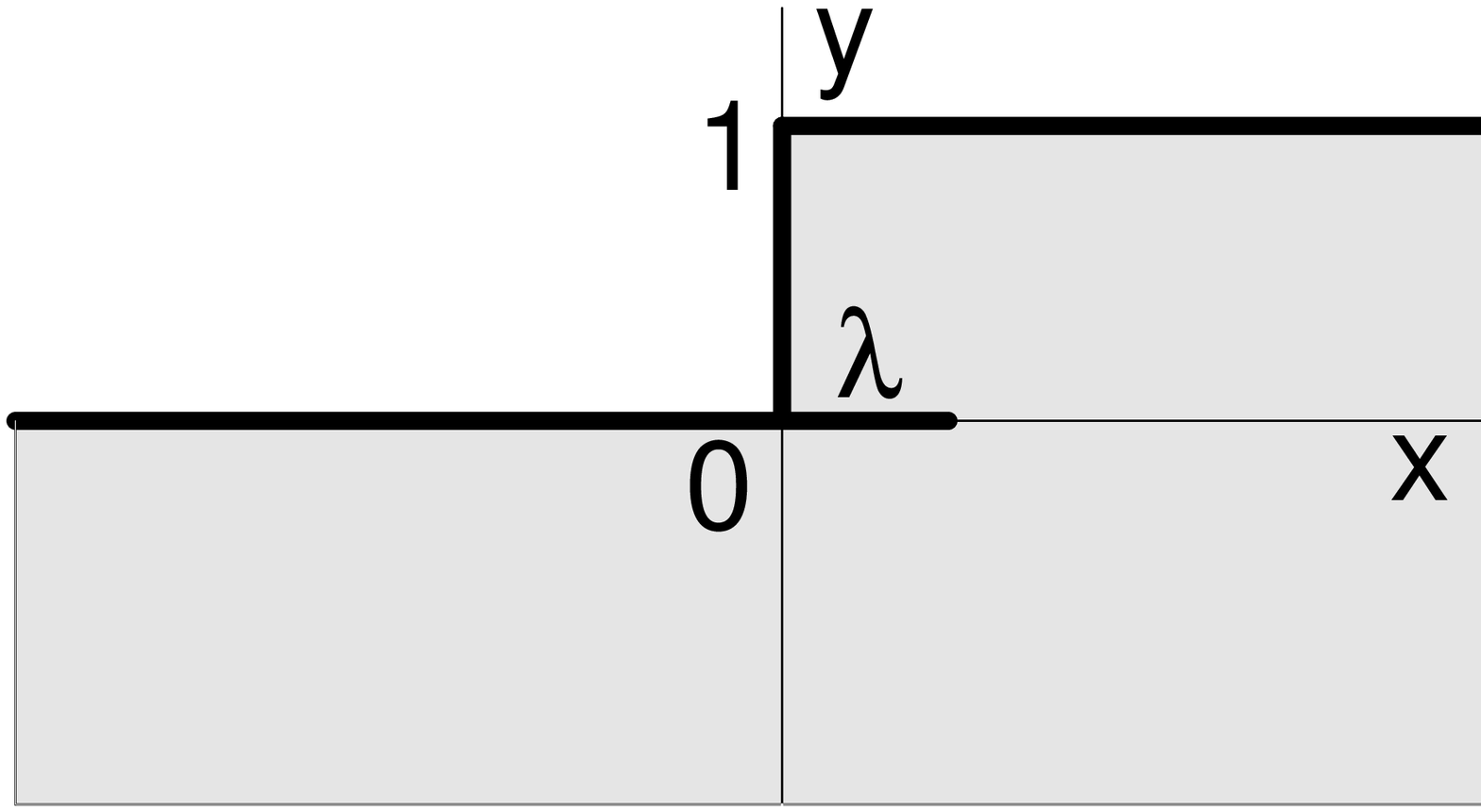, width=4cm,height=4cm}

\vskip -3.8cm
\hskip 4cm \epsfig{file=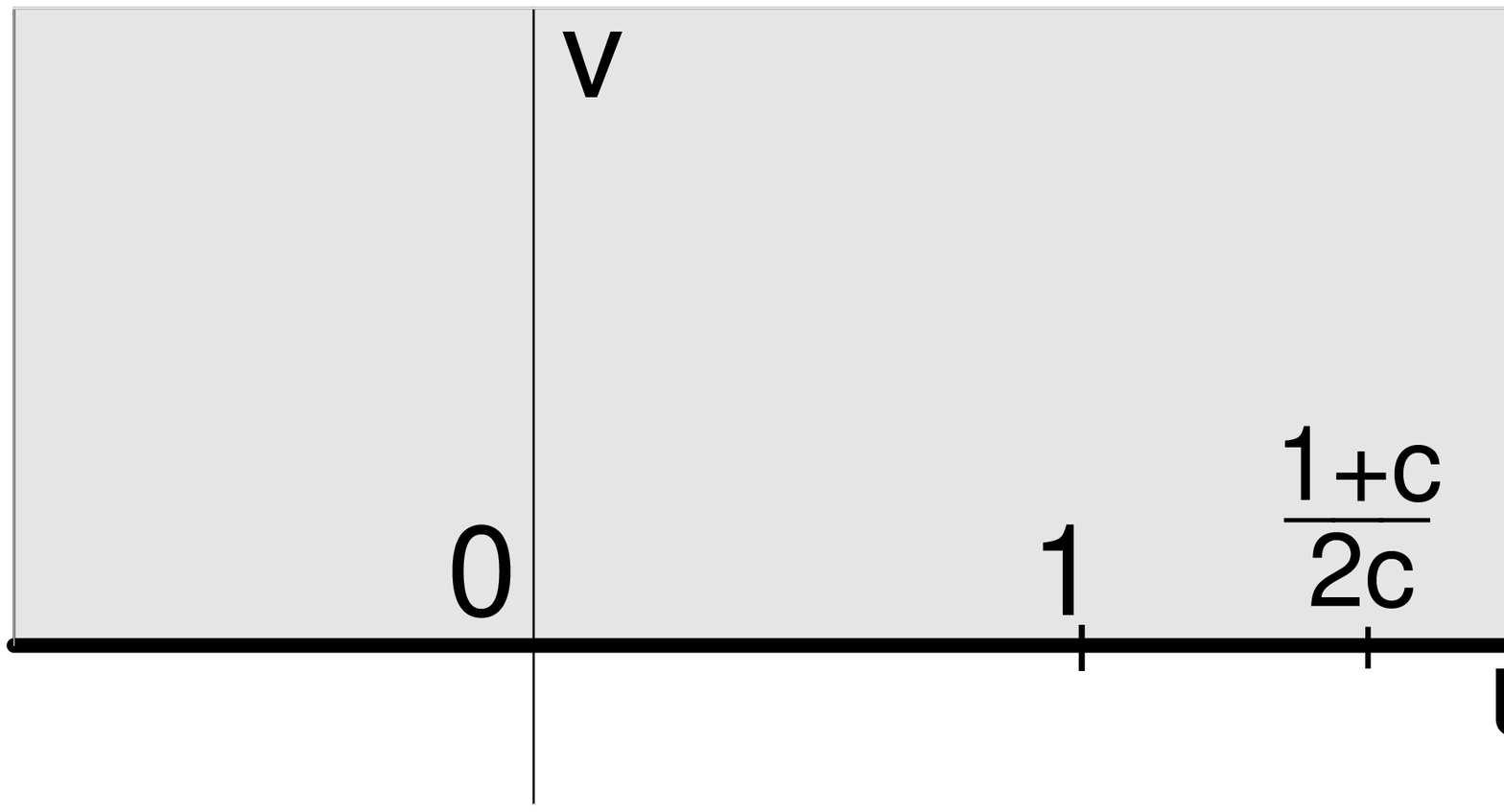, width=4cm,height=4cm}
\vskip -0.8cm
\noindent
{\small FIG. 2(a). Plane $z=x+iy$}~~~~~~{\small
FIG. 2(b). Plane 
$w=u+iv$}
\vskip 0.3cm
Our Dirichlet problem with the BC on the real axis $\Im w =0$ (which
come from the previous section),$$ 
\Phi (u,0)=\cases{1,&if $-\infty<u\leq 0$\cr
		  \phi (a,y(u)),&if $0<u<1$\cr
		  0,&if $u\geq 1$,\cr}\eqno(15),$$
has the solution $$
\Phi (u,v)={v\over \pi}\int_{-\infty}^\infty{\Phi (s,0)\over
(u-s )^2+v^2}ds\eqno(16)$$
in the upper half-plane $w$. Here 
by (14) $y(u)=2\pi^{-1}[\arccos{\sqrt{u}}-c\sqrt{u(1-u)}].$
The potential $\Phi (u,v)$ expressed in variables $x,y$ represents 
$\phi (x,y)$ outside the space charge zone. Our task now is  
to match the inside electric field ${\partial \phi\over \partial x}(a,y(u))$
in the interval $0<u<1$ with the field outside $$
{\partial \Phi\over \partial x}(u,v=0)={\pi\sqrt{u(1-u)}\over 1+c(1-2u)}
{\partial \Phi\over \partial v}(u,0).\eqno(17)$$

{\it Continuity of the electric field}. The matching condition (4) 
guarantees continuity of the electric field at
the boundary between the space charge region Q with the vacuum.
Using (7) and (9) we have at $x=a$ inside the space charge region
the field intensity, 
\vskip-0.3cm
$${\partial \phi\over \partial x}=y^{4/3}\sum_l q_lk_l(1-e^{-2k_la})u_l(y)
\eqno(18),$$
\vskip-0.2cm\noindent 
which should be equal to the vacuum field (17).
The exponentially small terms $e^{-2k_la}$ can be dropped 
as $a$ is assumed large.
Both terms ${\partial \phi\over \partial x}$ and 
${\partial \Phi\over \partial x}$ depend on all parameters $q_l$, 
but in a different way. One cannot expect an exact equality because of
the approximations made. We rewrite the matching condition (4) as$$
G[y^{4/3}]+\sum_lq_l\{G[y^{4/3}u_l(y)]-k_ly^{4/3}u_l(y)\}\approx
0,\eqno(19)$$
\vskip-0.2cm\noindent
where the functionals $G$ can be written explicitly as integrals with a 
logarithmic singularity.

We minimize the least square divergence from zero of the expression
(19) on the interval $0.15<y<0.85$ without approaching the endpoints 
where our treatment is not entirely adequate. A  
standard procedure yields a set of linear algebraic equations 
for $q_l$. We did not go further than
$l_{max}=10$. After the $q_l$ are computed one can find the current 
density (6) and the parameter $\alpha$ (13).

The accuracy of this method can be evaluated to some degree by 
determining the relative average discrepancy $\Delta$ of electric fields at 
the boundary of the space charge region Q 
on the chosen interval of $y$. The results of our computations are shown
in the Table 1, where for different values of the ledge $\lambda$
one can see also $\alpha (\lambda)$, parameters $q_l$, $\mu_{max}$ 
(near $y=0$ and $x=\pm a$), 
and the relative height (see Fig. 3) of the current wings $j_{max}$ at 
$x=\pm a$. When we extend the interval of matching
the electric fields up to $(0.01,0.99)$ the
quantities in the table stay approximately the same, only $\Delta$
increases. This confirms the general validity of our method and
simultaneously 
 
shows that the computation of electric fields near the corners of the
rectangle Q is not very good. In particular, in the worst case (the 
most severe cathode regime, see also [6,7,12]) when 
$\lambda =0$, the electric field is singular at the cathode edges. The 
computation becomes unstable, we cannot therefore the 
data of Table 1 to be accurate there when the linearization fails too.

{\small $$\left |{\matrix{\lambda& 0&0.1&0.3&0.5&1&\infty\cr
\alpha&.6487&.5311&.3463&.2665&.2067&.1905\cr
\mu_{max}&1.955&1.432&0.804&0.605&0.497&0.461\cr
j_{max}&3.597&2.902&2.068&1.804&1.661&1.612\cr
\Delta&.0121&.0055&.0037&.0028&.0059&.0044\cr
q_{1}&.2448&.2339&.1891&.1530&.1140&.1032\cr
q_{2}&.2225&.1743&.0926&.0616&.0465&.0443\cr
q_{3}&.1867&.1411&.0720&.0528&.0448&.0433\cr
q_{4}&.1525&.0969&.0389&.0280&.0260&.0246\cr
q_{5}&.1184&.0760&.0327&.0253&.0232&.0222\cr
q_{6}&.0914&.0500&.0192&.0143&.0148&.0134\cr
q_{7}&.0595&.0342&.0142&.0110&.0109&.0100\cr
q_{8}&.0439&.0216&.0087&.0061&.0072&.0061\cr
q_{9}&.0203&.0108&.0046&.0033&.0037&.0032\cr
q_{10}&.0142&.0064&.0028&.0017&.0025&.0019\cr}}\right |$$}
\centerline {\small TABLE 1}
\vskip0.3cm

When $l$ runs from $1$ to $10$ the values of $q_l$
decrease approximately by a factor 20-50. Therefore $\alpha$ is evaluated
very well by (13) where the $k_l$ increase from $\sim 4$
to $32$. The accuracy of $j(x)$ and $\mu(x)$ 
might be improved if one truncates (9) at a larger $l_{max}$ though
we would not expect dramatic changes. The cases $\lambda =\infty$
(when the mapping is exact) and $\lambda =1$ are very close which 
means that the parameter
$\alpha(a,\lambda)$ as well as $q_l$ are approximately independent of
$a$ when $a>1$. Keeping the exponential terms in $\mu(x,y)$ in the 
matching conditions (20) does not complicate the calculation and it
will give 
 
only insignificant corrections. The current density $j(x)$ at $x=0$
increases in this case by about $2q_1e^{-3.9a}$. If $a$ is smaller,
but $2\lambda +2a>1$, the scheme of computation is the same though
the $q_l$ become functions of $a$ and one cannot decrease $a$ too much
because the first term in (9) needs corrections.

An important part of our analysis is the form
(11) of $f_l(x)$ which implies that the current density (10) in a narrow 
region of width $\sim 1$ ($D$ in the original units) at the cathode edges
has a sharp peak which decays faster than $\exp [-3.88(a-|x|)]$.
Everywhere else $j(x)$ is close to the 1-d current $j_1(x)=1$ with the 
exponentially small corrections. {}For illustration plots of the
current density distribution (6) are shown in Fig.3 for different
widths of the cathode. 
\vskip -2.2cm
\hskip -0.6cm \epsfig{file=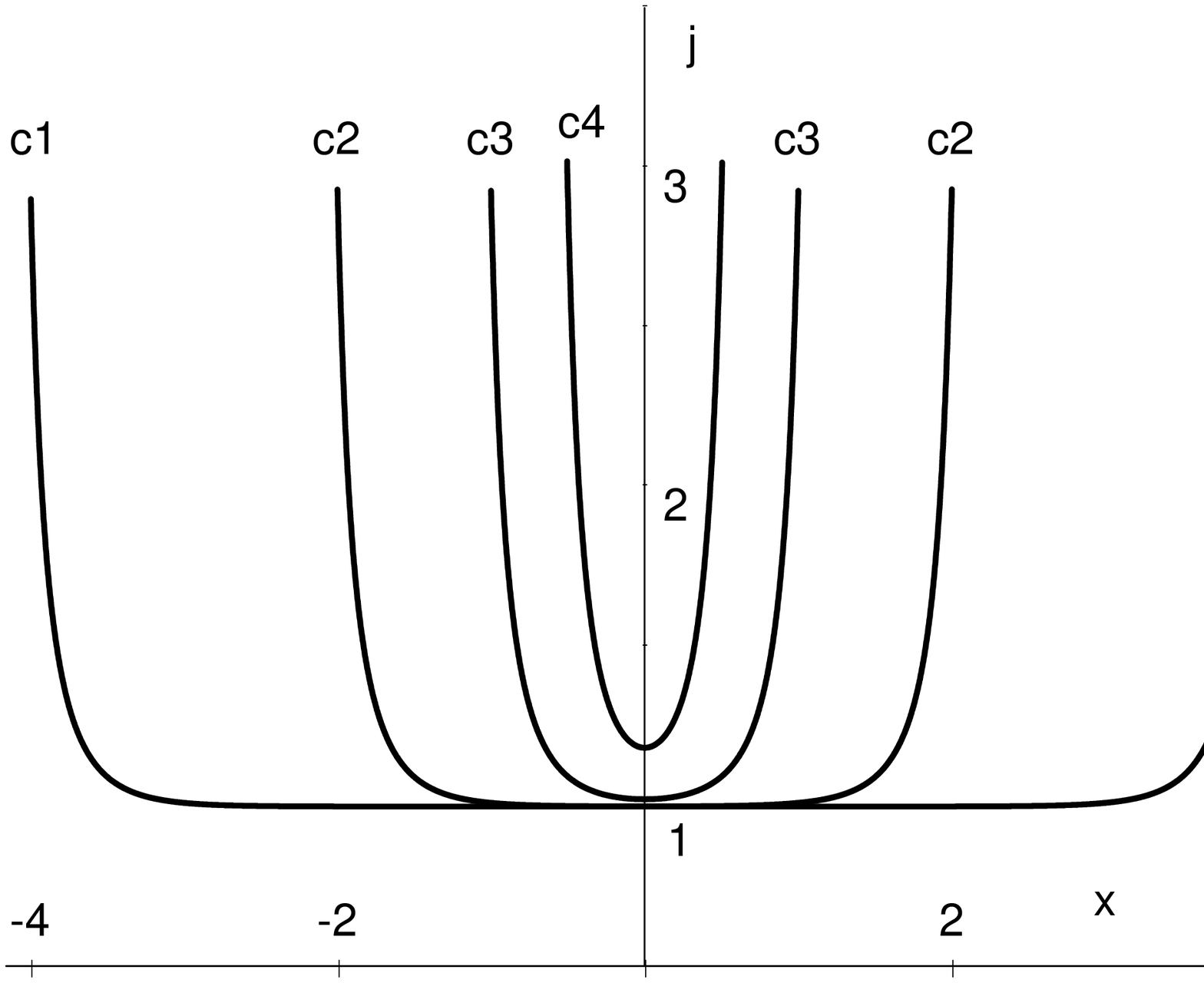, width=7cm,height=7cm}
\vskip 0cm

{\small FIG. 3. The current densities when $\lambda =0.1$,
$c1-c1,\ c2-c2,\ c3-c3$ and $c4$ correspond to
$2a=8,4,2,1$ respectively.}
\vskip 0.4cm
We can compare our curve $c3$ for $a=1$ with PIC simulations
presented in [7]. There for zero cathode recess ($dx=0$), $A=50$ mm,
and unfortunately unspecified width of the shroud
a reasonable fit would be
$j_{max}=3.9$ in our Table for $\lambda =0.1$
versus $3.2$ in [7]. We get the half-width of the current density 
peak  $\sim 1.2$ mm while in [7] it was $1$ mm. Our magnetic field is 
stronger and we think also that PIC simulations with finer grids 
are closer to our computation, but diverge from 
experimental results because the real cathodes with their finite
thickness and roundness do not have the very strong electric field 
intensities present however in the models.

{\it Generalization}. We expect that this pattern of narrow wings of
the current density holds also for finite
flat cathodes with perimeter $P$ where 
the boundary region will have an area $\sim PD$ if we assume
reasonable restrictions on the curvature and return
back to the original units. In the general case of a
cathode with area $S\gg PD$ the total current can be written as 
the sum $I=SJ_1+ PD {\tilde J}$. The "edge" current ${\tilde J}$,
which is assumed here to be independent of geometry, can be viewed as  
distributed over the edges of the cathode of width equal to the 
distance between the electrodes. The ratio ${\tilde J}/J_1$ can be 
evaluated in terms of the parameter $\alpha$ defined in (13). 
Comparing  
$\bar J=I/S=J_1(1+{\tilde J}PD/J_1S)$ with eq.(12) for our geometry, where 
$\bar J=J_2,\ S=2AL$ and $P=2L$ (the length $L$ of the cathode segment
is arbitrary), we have ${\tilde J}=J_1\alpha /2$ and finally
\vskip-0.5cm
$$\bar J=J_1\left (1+\alpha {PD\over 2S}\right ),\eqno(20)$$
\vskip-0.2cm\noindent
which should be applicable in general situations. In particular
the factor $PD/2S$ in (20) becomes $D/R$ for a circular cathode of 
the radius $R$ and $2DE(\sqrt{1-C^2/B^2})/\pi C$ for an elliptical 
cathode with the half-axes $B>C$, where $E(k)$ is the complete
elliptical integral. {}For a rectangular cathode with the sides $L$ and 
$H$ it is equal to $D(L^{-1}+H^{-1})$.

{\it Conclusions.} 1) The current wings, Fig.3, resemble 
simulated ones [6,7,12]. They are high when the width of
ledges $\lambda$ is small and the vacuum electric field near the 
cathode edges is strong. Their form becomes practicaly constant when 
the ledges are wider than the distance D between electrodes. 2) The
shape of the current wings, which is determined by eigenvalues
$k_l$ of (8), is roughly exponential and the 1-d current is 
restored up to a few percents at the distance D from the edges.
3) The parameter $\alpha$, which defines the net current density,  
depends on the width of ledges. An approximate empirical formula 
\vskip-0.3cm
$$\alpha(\lambda)\approx 0.19 +0.48e^{-3.7\lambda},\eqno(21)$$
\vskip-0.1cm\noindent
agrees with the data in Table 1 within $\sim 3.3\%$. (For a different
model with the constant current density $\alpha$ was estimated in [8]
as close to $0.31$.) 4) Our techniques of matching the electric fields at the
boundary of the space charge region and using rather modest variations
of the potential in the $x$ direction is effective for approximate 
modelling the 2-d and 3-d flows of charged particles.  

\bigskip
{\it Acknoledgements}. We thank R.Barker, R.J.Umstattd, and 
O.Costin for inspiration and useful comments. Research supported by 
AFOSR Grant \# F49620-01-0154.


\vskip-0.6cm

\begin{thebibliography}{12}

\bibitem{[1]} C.D.Child, Phys. Rev. {\bf 32}, 492 (1911); I.Langmuir,
Phys. Rev. {\bf 2}, 450 (1913).

\bibitem{[2]} I.Langmuir and K.B.Blodgett, Phys. Rev. {\bf 22}, 347
(1923); I.Langmuir and K.B.Blodgett, Phys. Rev. {\bf 24}, 49 (1924).

\bibitem{[3]}  D.C.Barnes and R.A.Nebel, Phys. Plasmas {\bf 5}, 
2498 (1998); R.A.Nebel and D.C.Barnes, Fusion Technology {\bf 38}, 28 (1998).

\bibitem{[4]} A.S.Gilmour, Jr., {\sl Microwave Tubes} (Artech House, Dedham,
MA, 1986); P.T.Kirstein, G.S.Kino, and W.E.Waters, {\sl Space Charge
Flow} (McGraw-Hill, New York, 1967); A.Valfells, D.W.Feldman, M.Virgo, 
P.G.O'Shea, and Y.Y.Lau, Phys. Plasmas {\bf 9}, 2377 (2002).

\bibitem{[5]} J.W.Luginsland, Y.Y.Lau, R.J.Umstattd, and J.J.Watrous,
Phys. Plasmas {\bf 9}, 2371 (2002). 

\bibitem{[6]} R.J.Umstattd and J.W.Luginsland, Phys. Rev. Lett. {\bf 87}, 
145002 (2001)

\bibitem{[7]} F.Hegeler, M.Friedman, M.C.Myers, J.D.Sethian, and S.B.
Swanekamp, Phys. Plasmas {\bf 9}, 4309 (2002).

\bibitem{[8]} J.W.Luginsland, Y.Y.Lau, and R.M.Gilgenbach,
Phys. Rev. Lett. {\bf 77}, 4668 (1996); Y.Y.Lau, Phys. Rev. Lett. 
{\bf 87}, 278301 (2001).

\bibitem{[9]} Y.Y.Lau, P.J.Christenson, and D.Chernin, Physics of 
Fluids B{\bf 5}, 4486 (1993).

\bibitem{[10]} A.Erdelyi (editor), {\sl Higher Transcendental Functions} 
Vol. 2 (McGraw-Hill, New York, 1953).

\bibitem{[11]} W.von Koppenfelds and F.Stallmann, {\sl Praxis der
Konformen Abbildung} (Springer-Verlag, Berlin, 1959).

\bibitem{[12]}  R.J.Umstattd, D.A.Shiffler, C.A.Baca,
K.J.Hendricks, T.A.Spencer, and J.W.Luginsland, Proc. SPIE
Int. Soc. Opt. Eng. {\bf 4031}, 185 (2000).

\end{thebibliography}
\end{document}